\documentstyle[12pt]{article}

\def\H{{\cal H}}

\newcommand{\be}{\begin{eqnarray}}
\newcommand{\en}{\end{eqnarray}}

\begin{document}
\begin{titlepage}
\begin{flushright}
EFI 96-42\\
MPI-Ph/96-82  \\
hep-th/yymmnn
\end{flushright}

\begin{center}  
\vskip 0.3truein

{\Large\bf {Duality, Superconvergence }} \\
\vskip0.07truein
{\Large\bf {and the Phases of Gauge Theories}}

\vskip0.5truein

{Reinhard Oehme} 
\footnote{E-mail: oehme@control.uchicago.edu}

\vskip0.2truein

{\it Enrico Fermi Institute and Department of Physics}

{\it University of Chicago, Chicago, Illinois, 60637, USA}
\footnote{Permanent Address}

{\it and}

{\it Max-Planck-Institut f\"{u}r Physik}

{\it - Werner-Heisenberg-Institut -}

{\it 80805 Munich, Germany}
\end{center}
\vskip0.2truein \centerline{\bf Abstract} \vskip0.13truein

Results about the phase structure of certain $N=1$ supersymmetric
gauge theories, which have been obtained as a consequence of
holomorphy and `electric-magnetic' duality, are shown to be in {\it
quantitative} agreement with corresponding consequences of analyticity 
and superconvergence of the gauge field propagator. This connection is 
of interest, because the superconvergence arguments for confinement are 
{\it not} restricted to theories with supersymmetry. The method of 
reduction in the space of coupling parameters is used in order to define,
beyond the matching conditions, an asymptotically free,
dual magnetic theory involving Yukawa couplings.

\end{titlepage}
\newpage
\baselineskip 20 pt
\pagestyle{plain}


\setcounter{equation}{0}

\vskip0.2truein

The phase structure of supersymmetric gauge theories has recently been
elucidated with the help of holomorphy and duality
\cite{WIS,SEI,SEN}. These features seem to be characteristic for SUSY
theories, where they provide insight into the non-perturbative
structure. It is an important question, to what extent the results
obtained for SUSY theories are generic, and have parallels in ordinary
gauge theories like Quantum Chromodynamics.

We are mainly considering theories with the gauge groups $SU(N_C)$ or
$SO(N_C)$, and with $N_F$ flavors of zero mass matter fields.  For
SUSY theories, duality arguments lead to confinement for values of the
number of flavors $N_F$ which do not reach up to the point where
asymptotic freedom is lost.  There is a region in $N_F$, where the
theories are in an interacting non-Abelian Coulomb phase. For these
values of $N_F$, there is no confinement, neither of the `electric'
nor of the `magnetic' excitations of the theory.

Some time ago, we have developed arguments for the confinement of
gluons and quarks in theories like QCD \cite{ROC, NIC}.  These
arguments are based upon superconvergence relations of the gauge field
propagator \cite{WZS, ROS, OWG}. There are two approaches. One is more
heuristic and considers the potential between static color charges
\cite{RLP, NIP}, the other is more formal and subtile, involving the
definition of the physical state space $\H$ in terms of the BRST
algebra \cite{ROC}. Both methods give confinement for values of $N_F$
below an upper limit, which is lower than the value where the one-loop
$\beta$ function coefficient vanishes and asymptotic freedom is
lost. We have a finite region with asymptotic freedom and no
confinement.

Already in \cite{LOP}, we have applied the superconvergence arguments
to SUSY theories in the Wess-Zumino representation. Hence we can now
compare with the predictions of the new duality analysis. It is the
main purpose of this note to show that there is quantitative
agreement between both approaches, and to further explore the r\^{o}le
of analytic and asymptotic properties of the propagator for the
problem of confinement.
\footnote{A preliminary report about these results has been presented
at the International Workshop on High Energy Physics, Novi Svit,
Crimea, September 1995 \cite{NVS}.}  Since the superconvergence
arguments are valid for SUSY and non-SUSY theories, the comparison
underlines the generic character of the obtained phase structure.

\vskip0.2truein

An essential quantity in our superconvergence arguments is the ratio
$\gamma_{00}/\beta_0$, where $\gamma_{00}$ is the one-loop anomalous
dimension coefficient in the Landau gauge $\alpha=0$, and $\beta_0$ is
the one loop $\beta$-function coefficient.  The asymptotic expansions
\be \gamma(g^2,\alpha) &=& (\gamma_{00} + \alpha \gamma_{01} ) g^2 ~+~
\cdots ~, \cr \beta(g^2) &=& \beta_0 g^4 ~+~ \cdots ~
\label{1}
\en give the limits $g^2 \rightarrow 0$ of the anomalous dimension and
the renormalization group function.  From analyticity and
renormalization group properties of the gauge field structure function
$D (k^2,\kappa^2,g,\alpha)$, we find that it is the ratio
$\gamma_{00}/\beta_0$, which essentially determines the asymptotic
limit {\it in all covariant gauges} ($\alpha \geq 0$) and in {\it all
directions} of the complex $k^2$-plane.  Here $\kappa^2 < 0 $ is the
normalization point, and we have assumed that $\beta_0 < 0$. For the
discontinuity of the function $D$ along the positive, real $k^2$-axis,
we obtain
\begin{eqnarray}
-k^2 \rho (k^2,\kappa^2,g,\alpha) & \simeq& \frac{\gamma_{0
0}}{\beta_0} C_R (g^2,\alpha) \left(-\beta_0 \ln \frac{k^2}{\vert
\kappa^2\vert}\right)^{-\gamma_{0 0}/\beta_0 - 1} + \cdots ~.
\label{2}
\end{eqnarray}
The appearance of the coefficient $\gamma_{00}$ is due to the fact
that the gauge parameter $\alpha$ is renormalized, and $\alpha=0$ is a
UV-fixed point.  For this reason, $\gamma_{00}$ can be of physical
relevance, similar to the coefficient $\beta_0$.  In fact, for $N=1$
SUSY theories, we will show in the following that $2\gamma_{00}$ is
given by the negative of $\beta^d_0$, the one-loop $\beta$-function
coefficient of the dual theory.

In the following we work in the Landau gauge.  Although we have the
asymptotic properties of propagators in all covariant gauges
\cite{OWG}, the renormalization group equations are most easy to
handle in the Landau gauge. In any case, since confinement is a
physical aspect of the theory, it is easy to show that it is
sufficient to argue in a particular gauge.  We have the
superconvergence relation \cite{WZS,ROS} 
\be 
\int_{-0}^{\infty}d k^2
\rho (k^2, \kappa^2, g ) {}~~= ~~0 ~,
\label{3}
\en 
which is valid provided the ratio $\gamma_{00}/\beta_0 > 0$ , and
hence for $\gamma_{00} < 0$ in the presence of asymptotic
freedom. This relation gives a connection between high- and low-energy
properties of the theory.

In \cite{ROC} we have described in detail how one can make use of the
superconvergence relation in order to show, that states involving
transverse gauge field excitations are not elements of the physical
state space $\H$, as defined using the BRST algebra. Together with
other unphysical states of the theory, they form quartet
representations of the algebra, whereas physical states should be
singlets. In the more heuristic approach to confinement \cite{RLP}, we
use a dipole representation of the structure function $D(k^2) = 
\int_{-0}^{\infty}d k^{\prime2} \sigma(k^{\prime2}) (k^{\prime2} -
k^2)^{-2} $, with
the weight function $\sigma (k^2)~=~ \int_{-0}^{k^2} d {k'}^2
\rho({k'}^2) ~~$.  For $\gamma_{00}/\beta_0 > 0$, we have $\sigma
(\infty ) = 0~, ~~\sigma (k^2) > 0~, ~~{\sigma}' (k^2) = \rho (k^2) <
0 {}~$ for sufficiently large values of $k^2$. Together with
$C_R(g^2)>0$, we use these properties in order to argue for an
approximately linear confining potential.  There is no indication of
such a potential if $\gamma_{00}/\beta_0 < 0$. In this case, the 
dipole representation is still valid, but $\sigma(k^2) \rightarrow 
\infty$ for $k^2 \rightarrow \infty$.

If we consider theories with a gauge group like $SU(N_C)$, with
massless matter fields in the fundamental representation, the
coefficient $\gamma_{00}$ generally has a zero as a function of $N_F$
which is below the point where $\beta_0$ vanishes. For example, let us
take $SU(N_C)$ gauge theory with the matter $N_F \times ({\bf N_C} +
{\bf {\overline {N}}_C})$. Then 
\be 
\frac{\gamma_{00}}{\beta_0} ~~=~~
\frac{-\frac{13}{6}N_C ~+~\frac{2}{3}N_F} {-\frac{11}{3}N_C ~+~
\frac{2}{3}N_F}.
\label{4}  
\en

We have superconvergence, and hence confinement, for $N_F <
\frac{13}{4}N_C$, because $\gamma_{00}/\beta_0 > 0$. Furthermore,
there is an interval 
\be 
\frac{13}{4}N_C ~<~ N_F ~<~ \frac{22}{4}N_C
~,
\label{5}
\en 
where $\gamma_{00}/\beta_0 < 0$. In this region, superconvergence
is lost, but not asymptotic freedom. Our arguments for confinement do
not apply, and the study of the potential suggests that there is no
confinement. There must be a phase transition around the point
$N_F=\frac{13}{4}N_C$ where $\gamma_{00}$ changes sign, and where we
still have $\beta_0<0$.

In the article \cite{LOP}, we have applied our arguments for
confinement to SUSY theories. The appropriate quantity is the gauge
field propagator in the Wess-Zumino gauge, in particular if we ask the
question whether the elementary, transverse gauge field excitations
are excluded from the physical state space $\H$.  For $N = 1$
supersymmetric gauge theories, we write the one-loop coefficients of
the function $\beta (g^2)$ and the anomalous dimension $\gamma (g^2,
0)$ in the form 
\be \beta_0 ~~=~~ (16\pi^2)^{-1} \left( -3
C_2 (G) ~~+~~\sum_{i} n_i T(R_i) \right) ~~,
\label{6}
\en and \be \gamma_{00} ~~=~~ (16\pi^2)^{-1} \left( -\frac{3}{2} C_2
(G) ~~+~~ \sum_{i} n_i T(R_i) \right) ~~,
\label{7}
\en where $n_i$ is the number of $N = 1$ chiral superfields in the
representation $R_i$. These coefficients are determined by the
elementary field content of the theory. We have \be \gamma_{00} ~~<~~
0~~, ~~\beta_0 ~~<~~ 0 ~~~ for \cr \sum_{i} n_i T(R_i) ~~ <~~
\frac{3}{2} ~ C_2 (G) ~~,
\label{8}
\en which is the condition for the validity of the supercovergence
relation (\ref{3}) for the structure function, and hence for the
confinement of the elementary transverse gauge field excitations.  For
$G = SU(N_C)$, and matter fields in the fundamental representation
$N_F \times ({\bf N_C} + {\bf {\overline {N}}_C})$, we have
$\gamma_{00} < 0 $, $\beta_0 < 0 $ for $N_F < \frac{3}{2} N_C $, where
$N_F$ refers to four-component spinors in contrast to $n_i$. As
mentioned before, the superconvergence argument implies confinement
for $N_F < \frac{3}{2} N_C $, and we have indications of a
de-confining phase transition
\footnote{See page 450 of \cite{LOP}, where the existence of a phase
transition at $N_F = \frac{3}{2}N_C$ has already been pointed out, and
\cite{ROC} for the corresponding non-SUSY result.}  at $N_F =
\frac{3}{2} N_C$ , as $N_F$ increases \cite{LOP}. Above this
transition point, there is the region \be \frac{3}{2}N_C ~<~ N_F ~<~
3N_C ~~,
\label{9}
\en where there is no superconvergence, and hence no confinement.  It
corresponds to the interval given in Eq.(\ref{5}) for the non-SUSY
case.

\vskip0.2truein

After these preliminaries, we would like to connect the results
obtained using superconvergence, with the picture which emerges from
electric magnetic duality of $N=1$ SUSY gauge theories. As the
electric theory, we use again the gauge group $G=SU(N_C)$, with
massless matter fields in the representation $N_F \times ({\bf N_C} +
{\bf {\overline {N}}_C})$. For appropriate values of $N_C$ and $N_F$,
the corresponding dual magnetic theory has the gauge group $G_d=SU(N_F
- N_C)$ with $N^d_F=N_F$ flavors of magnetic chiral superfields and a
certain number of gauge-singlet massless superfields.

The one loop coefficients of both theories are given by: \be G =
SU(N_C) ~~~"electric"~~~ N = 1 ~~ SUSY \cr
{}~~~~~~~~~~~~~~~~~~~~~~~~~~~~~~~~~~~~~~~~~~~~\cr \beta_0
~~~=~~~(16\pi^2)^{-1} (-3N_C ~+~ N_F ) \cr \gamma_{00}
~~~=~~~(16\pi^2)^{-1} (-\frac{3}{2} N_C ~+~ N_F) ~~,
\label{10}
\en and \be G = SU(N_F-N_C) ~~~"magnetic"~~~N = 1 ~~ SUSY \cr
{}~~~~~~~~~~~~~~~~~~~~~~~~~~~~~~~~~~~~~~~~~~~~~~~\cr \beta^d_{0}
~~~=~~~(16\pi^2)^{-1} (-2N_F ~+~ 3N_C) \cr \gamma^d_{00}
~~~=~~~(16\pi^2)^{-1} (-\frac{1}{2} N_F ~+~ \frac{3}{2} N_C ) ~.
\label{11}
\en Here the coefficients of the dual map have been evaluated at the
same number of flavors $N^d_F = N_F $ as the original, electric
theory. However, these flavors refer to representations of the
magnetic gauge group.

{}From the above equations, we can extract the following important
relationships between electric and magnetic coefficients \cite{NVS}:
\be \beta^d_0 (N_F) ~&=&~ - 2 \gamma_{00} (N_F) ~~,
\label{12}
\en \be \beta_0 (N_F) ~&=&~ - 2 \gamma^d_{00} (N_F) ~~,
\label{13}
\en where it is again understood, that the variable $N_F$ on both
sides refers to matter fields with different quantum numbers in the
electric and magnetic functions respectively.  The appearance of the
factor `two' in Eqs. (\ref {12}, \ref{13}) is due to our definition of
the anomalous dimension by $u\frac{\partial R}{\partial u} = \gamma
R$, where $R = -k^2 D(k^2)$ and $u = k^2/\kappa^2$.

The duality relationships (\ref{12}, {13}) are not restricted to the particular
model considered. For example, $N=1$ supersymmetric
gauge theory with the group $G = SO(N_C)$ and with $N_F$
flavors in the representation $N_F \times {\bf N_C}$, has a dual map
with the group $G^d = SO(N_F-N_C+4)$ \cite{IOS}. The duality relations are
again given by equations (\ref{12}) and (\ref{13}). For this supersymmetric
theory with $G = SO(N_C)$, the coefficient $\gamma_{00}(N_F)$ changes
sign at $N_F = \frac{3}{2} (N_C - 2)$. It is certainly of
interest to study further models. We plan to discuss $SO(N_C)$ and 
other gauge theories elsewhere.

In writing the one-loop coefficients for the magnetic theory, we see
that there is no contribution from the Yukawa coupling of the 
singlet meson fields $M^i_j $ with the $N_F$ flavors 
of magnetic quark fields $q_i$ and ${\overline{q}}^j$. 
The corresponding superpotential is of the form
$\sqrt{\lambda} M^i_j q_i{\overline{q}}^j $  \cite{SEN, IOS}.
By itself, the Yukawa coupling $\lambda$
is not asymptotically free. As the dual theory, however,  
we should consider one which is obtained by
the method of the `reduction of couplings' \cite{OZS}, 
\cite{NPC}.  With this method, the
Yukawa coupling is expressed as a function of the gauge coupling,
$\lambda = \lambda(g^2)$, so that the resulting theory depends upon the
gauge coupling only,  satisfies the appropriate renormalization
group equations in this coupling, and preserves supersymmetry. As a
solution of the reduction equations, we get a theory which is 
UV-asymptotically free above $N_F = \frac{3}{2}N_C$, or IR-free 
for $N_F$ in an interval below the transition point, 
with $\lambda(g^2)$ being proportional to $g^2$ 
for $g^2 \rightarrow 0$ . The Green's functions of the reduced theory
have asymptotic power series expansions in $g^2$ in this limit.
We see that the one-loop
coefficient $\beta^d$  is not affected by the Yukawa coupling,
nor is the coefficient $\gamma^d_{00}$ in view of the singlet
character of the $M$ fields.  For our results about the de-confining
phase transition, we need only Eq. (\ref{12}).

To be more specific about the construction of the dual map, using
reduction in coupling parameter space, we briefly describe some 
aspects of the procedure for the particular system considered here.
\footnote{For a recent survey of the reduction method, see
\cite{PRO}. The case of two couplings has been discussed in
detail in \cite{OZC}.} 
The result is also relevant for the infrared fixed point to be mentioned later.

For the gauge group, we take a generic $G = SU(N)$, where 
$N = N_F - N_C$ in the case of the
magnetic theory described above. For the following
discussion, we omit the label $d$ indicating the dual system.
In the usual renormalization group equations for the effective couplings
$g^2(u)$ and $\lambda(u)$, with $u=k^2/\kappa^2$, we eliminate the scaling
variable $u$, and obtain the reduction equations
\begin{eqnarray}
\beta (g^2, \lambda(g^2)) \frac{d \lambda(g^2)}{d g^2} ~=~ 
\beta_\lambda (g^2, \lambda(g^2))~,
\label{14}   
\end{eqnarray} 
This is an equation 
for the Yukawa coupling $\lambda$ as a function of $g^2$. The
$\beta$-functions for the original two-parameter theory
are assumed to have asymptotic power series
expansions. General two-loop results for SUSY theories may
be found in the references \cite{AND}.
For the model considered, the $\beta$-functions  are of the form 
\begin{eqnarray}
\beta (g^2, \lambda)~&=&~\beta_0~g^4~+~(\beta_1~g^6~+~ 
\beta_{1,\lambda}~g^4\lambda) ~+~\cdots  \cr
\beta_\lambda (g^2, \lambda)~&=&~c_\lambda g^2 \lambda ~+~ 
c_{\lambda\lambda} \lambda^2 ~+~\cdots ~.
\label{15}
\end{eqnarray}
The coefficients are given by
\begin{eqnarray}
\beta_1~&=&~(16\pi^2)^{-2} \left( 2N(-3N + N_F) ~+~ 
4N_F \frac{N^2 - 1}{2N} \right) \cr
\beta_{1,\lambda}~&=&~(16\pi^2)^{-2} \left( -2N^2_F \right)  \cr
c_{\lambda\lambda}~&=&~(16\pi^2)^{-1} \left( N + 2N_F \right) \cr
c_\lambda~&=&~(16\pi^2)^{-1} \left( -4 \frac{N^2 - 1}{2N} \right) ~.
\label{16}
\end{eqnarray}
With the expansions of the $\beta$-functions as given, we see
that the differential equation (\ref{14}) is singular at $g^2=0$.
By a uniformization transformation, we can remove the singularity,
and show that all solutions in the neighborhood of the origin
are given by asymptotic series expansions 
\cite{OZC, WZG, ONA}. They may contain
non-integer powers (and logarithms, in general). However, we are mainly
interested in {\it special} solutions \cite{OZC, ONA},
which have a power series
expansion. For the system considered, we write these solutions
in the form
\be
\lambda(g^2) = g^2 f(g^2)~, ~\mbox{with}~~ f(g^2) = f_0 + \sum_{m=1}^{\infty} 
\chi^{(m)} g^{2m} ~.
\label{17}
\en
Substitution into the reduction equation yields the fundamental relation 
\be
\beta_0 f_0 ~=~ \left( c_{\lambda\lambda} f_0 ~+~ 
c_\lambda \right) f_0 ~.
\label{18}
\en
There are two solutions:
\be
f_0 = f_{00} = 0~~ \mbox{and}  ~~f_0 = 
\frac{\beta_0 - c_\lambda}{c_{\lambda\lambda}} ~.
\label{19}
\en
For $\beta_0 < 0$, we find that, in both cases, the coefficients $\chi^{(m)}$ 
in Eq. (\ref{17}) are uniquely determined
by the expansions of the  $\beta$-functions. For the solution $f_{00}$,  
it follows directly that $\chi^{(m)} = 0$ for all $m$.
It corresponds to the $SU(N)$ theory without the Yukawa coupling.
Of main interest as a dual map is the second solution.
Using regular reparametrizations of the 
theory, we can remove the coefficients  $\chi^{(m)}$ 
for this solution. Then we have the reduced
theory with 
\be
\lambda(g^2)~=~g^2f_0 ~~\mbox{with}~~  
f_0~=~\frac{\beta_0 - c_\lambda}{c_{\lambda\lambda}} ~,
\label{20}
\en
modulo terms which vanish faster than any power. The solution
(\ref{20}) represents an asymptotically free theory, which depends
only upon the gauge coupling $g^2$, and contains no further arbitrary 
parameters. In order to use this solution as a physical theory,
we must require that the coefficient $f_0$ is positive. From Eqs.
(\ref{16}) and (\ref{20}) we see that $f_0 > 0$ provided
$N_F > N + 2/N$ for the $SU(N)$ theory with Yukawa coupling.
With $N = N_F - N_C$ , as required for the dual theory, the 
positivity condition is $N_F > N_C + 2/N_C$. Interestingly, it
is of the same form as for the $SU(N)$ theory.
It follows, that we can certainly use
this solution in the window $\frac{3}{2}N_C < N_F < 3N_C$ and above; only for
$N_C = 2$ does the bound touch the lower end of the window.

It remains to discuss the {\it general} solution of the reduction
equation (\ref{14}) for the case $\beta_0 < 0 $. In the models 
considered here, this solution
contains rational powers in the asymptotic expansion, with exponents
given by 
\be
\xi~=~\frac{\beta_0 - c_\lambda}{-\beta_0} ~
\label{21}
\en
and multiples thereof. Note that $\xi > 0$ in the window (\ref{9})
and above. It can also be written in the form
$\xi = \frac{c_{\lambda\lambda}}{-\beta_0}f_0$, where 
$c_{\lambda\lambda} > 0$. In all cases, even if $\xi$ should
happen to be an integer, the leading term in the asymptotic
expansion of the general solution is 
\be
\lambda(g^2) ~=~ A g^{2(1+\xi)} ~+~ \cdots ~,
\label{22}
\en
where $A$ is an undetermined coefficient. Once the factor $A$
is fixed, all higher order coefficients in the expansion are
determined. If expressed as a function of $N_F$ and $N_C$ for the
dual map with $SU(N_F - N_C)$, the exponent $\xi$ is given by
\be
\xi~=~\frac{N_C\left( N_F - N_C - {2}/{N_C} \right)}
{2(N_F - N_C)(N_F - \frac{3}{2}N_C)} ~,
\label{23}
\en
and it becomes large as $N_F$ approaches the lower end of the
window at $N_F = \frac{3}{2}N_C$. We see that the general solution
(\ref{22}) does not approach the special power series solution
(\ref{20}) in the limit $g^2 \rightarrow 0$. The latter is therefore
isolated or unstable with respect to it's embedding into the 
two-parameter theory. The ratio of both solutions does not approach 
`one'. This is a feature, which is quite common for SUSY theories
\cite{ONA, SUZ} .
Within the framework of theories with asymptotic freedom and
renormalized power series expansions, we consider the {\it special}
solution with $N = N_F - N_C$ as the appropriate 
dual of the original $SU(N_C)$ theory. As we have mentioned, 
the solution (\ref{20}) should be amended by non-perturbative
contributions, which vanish faster than any power in the limit
$g^2 \rightarrow 0$. These should be present, provided there are
corresponding additions to the $\beta$-functions.

So far, we have assumed that $\beta_0 < 0$. But the unique power
series solution (\ref{17}, \ref{20}) is also valid for $\beta_0 > 0$,
provided certain conditions are satisfied. One is the positivity 
condition $f_0 > 0$. It requires $N_C + 2/N_C < N_F < \frac{3}{2}N_C$,
an interval which is non-empty for $N_C > 2$. The other is,  that
the quantity $\xi$, as defined in (\ref{21}), 
is not a negative integer. (We refer to \cite{OZS} for the details
of the special case where $\xi = -n, ~ n = 1, 2, ...$).
Assuming that the conditions stated above are satisfied, we have a unique
solution for $\beta_0 > 0$. This {\it special} 
solution represents a renormalized,
IR-asymptotically free theory. As the dual map, it describes the 
low energy excitations of the system.

Given the conditions described above, 
the {\it general} solution for $\beta_0 > 0$  
has an asymptotic expansion for $g^2 \rightarrow 0$,
which can be brought into the form
$\lambda(g^2) \simeq f_0g^2 + Bg^{2(1+|\xi|)} + \cdots $,
with an undetermined coefficient $B$. For all $B$, this general
solution approaches the unique special solution (\ref{20}), 
which therefore is stable with respect to the embedding into
the two-parameter theory.

There remains the boundary case where $N_F = \frac{3}{2}N_C$,
and hence $\beta_0 = 0$ for the magnetic theory. In this 
situation, we have $f_0 > 0$ for $N_C > 2$. There is again
a unique power series solution of the form (\ref{17}, \ref{20}).

The reduction described above gives a more detailed picture
of the dual, magnetic theory, which is defined, a priori, on the 
basis of the matching conditions \cite{SEI, SEN}. It is 
important to have this picture for our comparison with the
results obtained on the basis of analyticity and superconvergence,
which uses the functions $\beta(g^2, \lambda(g^2))$ and
$\gamma(g^2, \lambda(g^2))$ of the reduced theory.
We note here, that it is of interest to consider the reduction 
equations away from the fixed point $g^2 = 0$, wherever one has some 
information about the $\beta$-functions \cite{ONA, STR}.
We hope to discuss the general solutions, and non-perturbative
aspects of the reduction, at another occasion.
 
\vskip0.2truein

Now we return to the phase transition at $N_F = \frac{3}{2}N_C$.
We revert to the notation where the label $d$ indicates the dual map.
It follows from Eqs. (\ref{12}, \ref{13}), that the zero of the one-loop 
anomalous dimension coefficient $\gamma_{00}(N_F)$,
which has emerged as a critical point in our superconvergence 
arguments for confinement, corresponds to a zero of the $\beta$-function
coefficient $\beta^d_0$ of the dual theory. There is an analogous
relationship between $\gamma^d_{00}$ and $\beta_0$. These connections
show, that the coefficient $\gamma_{00}$ is a characteristic quantity
for the structure of the gauge theory, on the same level as $\beta_0$.

Let us assume that $N_F \geq N_C + 2$, so that we have a non-Abelian
dual map of the $SU(N_C)$ theory. The point $N_F = N_C + 2$ is below
the critical value $N_F = \frac{3}{2}N_C$ for $N_C > 4$. In the 
region $N_C + 2 < N_F <  \frac{3}{2}N_C$ we have $\gamma_{00}/\beta_0 > 0$,
and the superconvergence arguments imply confinement of the
electric excitations, by showing that they are not elements of the physical state
space $\H$. According to Eq. (\ref{13}), the dual magnetic theory
has $\beta^d_0 > 0$ in this region of $N_F$. It is not asymptotically
free. Rather, we have non-interacting magnetic quanta in the 
infrared, which should be viewed as composites of the elementary
electric quanta. The magnetic theory looses asymptotic freedom
at $N_F = 3N^d_C = 3(N_C - N_F)$, which exactly corresponds to
the point $N_F =  \frac{3}{2}N_C$ where the coefficient $\gamma_{00}$
vanishes. For even smaller values of $N_F$,
like $N_F = N_C + 1$ and $N_F = N_C $,
the Higgs mechanism has removed the massless magnetic quanta of the dual
gauge theory. We have massive composites as physical states in
$\H$.

\vskip0.2truein

Of particular interest is the window 
$\frac{3}{2}N_C < N_F < 3N_C $ for the electric theory. 
With $N^d_F = N_F$, it exactly
corresponds to the region $ 3(N_F - N_C) > N_F > \frac{3}{2}(N_F - N_C)$
for the dual magnetic formulation. In this region. we have $\gamma_{00} > 0$
and $\beta_0 < 0$, and for the dual system, 
with our relations (\ref{12}, \ref{13}),
$\gamma^d_{00} > 0$ and $\beta^d_0 < 0$. Both, electric and magnetic versions,
are asymptotically free and have {\it no} supercovergence. 
Our superconvergence arguments indicates that there is no confinement.
The gauge field propagator is not compatible with an approximately
linear potential, and the BRST arguments do not prevent elementary
quanta from being elements of the physical state space.

In the window for $N_F$ discussed above, the the UV-asymptotic
behavior of the gauge field propagator for the electric
$SU(N_C)$ theory is given by
\begin{eqnarray}
-k^2 D (k^2)&\simeq& 
C_R \left(-\beta_0 \ln
\frac{k^2}{\kappa^2}\right)^{-\gamma_{00}/\beta_0} + \cdots ~,
\label{24}
\end{eqnarray}
which diverges since $\gamma_{00}/\beta_0 < 0$. Although this result
is valid in all covariant gauges ($\alpha \geq 0$),
we consider here only the Landau gauge for simplicity. For  
$N_F$ near the upper limit $3N_C$ of the region considered,
where $\beta_0$ approaches zero from below, the exponent in
Eq. (\ref{14}) is very large and and we have strong divergence. 
In contrast, for $N_F$ near the lower limit $\frac{2}{3}N_F$,
the coefficient $\gamma_{00}$ vanishes, and we have only small
modifications of an asymptotic $1/k^2$ behavior of the function
$D(k^2)$. As we continue to values of $N_F$ below the critical 
point $ \frac{3}{2}N_C$, where  $\gamma_{00}/\beta_0 > 0$,
the structure function is superconvergent, and we have confinement.
We see that the phase transition of the theory at $N_F = \frac{3}{2}N_C$   
is reflected in an essential change of the asymptotic behavior.

\vskip0.2truein

The UV-behavior described above for the $N=1$ SUSY theory with
$G=SU(N_C)$ in the window (\ref{9})
 is completely analogous to that of the corresponding
non-Susy gauge theory in the Coulomb phase comprising the interval
(\ref{5}).

\vskip0.2truein

Let us now consider the UV-limit of the gauge field structure
function for the dual theory with $G^d=SU(N_F-N_C)$. From
Eqs. (\ref{12},\ref{13}), we obtain the relation 
$\gamma^d_{00}/\beta^d_0 = \beta_0/4\gamma_{00}$, so that  
\begin{eqnarray}
-k^2 D^d (k^2)&\simeq& 
C^d_R  \left(\frac{1}{2}\gamma_{00}\ln
\frac{k^2}{\kappa^2_d}\right)^{-\beta_0/4\gamma_{00}} + \cdots ~.
\label{25}
\end{eqnarray}
The relation between $\kappa^2_d$ and $\kappa^2$ is given in 
\cite{IOS}.
As expected. the UV-behavior for the magnetic
theory in the Coulomb region is just the opposite
of the one described above for the electric case,
interchanging the upper and the lower end of the window (\ref{9}). 
The transition from the Coulomb to the confining phase is always associated 
with the loss of asymptotic freedom of the dual theory, the
excitation of which then give the observable low energy composites.

\vskip0.2truein

It has been argued, that the SUSY gauge theories we consider here,
have an IR-fixed point in the Coulomb interval (\ref{9})
\cite{SEN, FIX}. In the 
neighborhood of this fixed point $g_*^2$, we write the 
$\beta$-function of the magnetic theory in the form
\be
\beta^d(g^2) ~=~ \beta_*(g^2 - g_*^2) ~+~\cdots ~~,
\label{26}
\en  
where $\beta_* = \beta^{d\prime}(g^2_*) > 0$. Here $\beta^d(g^2) = \beta^d
(g^2, \lambda(g^2))$ is the renormalization group function of
the theory corresponding to the 
special solution of the reduction equation (\ref{14}) in the 
limit of small values of $g^2$. With
$\lambda(g^2)$ being a solution of this differential equation,
we see that $\beta^d (g^2_*) = 0$ implies $\beta^d_\lambda (g^2_*) = 
\beta^d_\lambda (g^2_*, \lambda (g^2_*)) = 0$, provided we require
that the solution is bounded at $g^2 = g^2_*$.

Assuming that the anomalous 
dimension $\gamma^d(g^2)$ is regular at $g^2 = g^2_*$, we find for the 
IR-limit of the gauge field structure function for $SU(N_F - N_C)$
\be
-k^2 D^d(k^2) ~\simeq~ C_* \left(\frac{k^2}{\kappa_d^2}\right)^{\gamma^d(g^2_*)} ~.
\label{27}
\en
The IR-fixed point $g^2_*$ is a function of $N_C$ and $N_F$. Using the 
asymptotic expansion (\ref{15}) in the special frame
where  $\lambda = g^2 f_0$, we can
obtain an approximate expression for $g^2_*$ in the limit where
$N_F$ approaches the lower end $N_F = \frac{3}{2}N_C$ of the window. Here
$g^2_*$ is proportional to $(N_F - \frac{3}{2}N_C)$, and hence vanishes in this limit.
With $\gamma^d(g^2_*) \simeq \gamma^d_{00}~g^2_* $ , and because $\gamma^d_{00} 
> 0$ in the window, we see that there is a slight softening of the 
$k^{-2}$-behavior of the structure function $D^d(k^2)$. 
The modificaton of the $k^{-2}$-behavior disappears at
the transition point $\frac{3}{2}N_C$, below which the theory looses asymptotic
freedom, and the IR-limit contains free magnetic gauge field excitations.
In a similar fashion, we can discuss the structure function of the
electric $SU(N_C)$ gauge theory at the upper end of the window near
$N_F = 3N_C$.

\vskip0.2truein

To sum up, for the SUSY model considered, criteria for confinement, which
are based upon analyticity and supercovergence of the gauge field propagator,
are in exact agreement with the results of the duality approach.
However, the superconvergence arguments are also applicable to non-SUSY
gauge theories like QCD, for which they were introduced originally.
The method of the reduction of coupling parameters has been used
in order to define the dual theory beyond the matching conditions. 
Even though it involves Yukawa couplings, in the window and above, 
the reduced magnetic theory is UV-asymptotically free, with a renormalized 
asymptotic power series expansion. In an interval below the window, it is
IR-free and describes the low energy excitations of the system.

\vskip0.5truein
\centerline{ACKNOWLEDGMENTS}

For conversations or communications.  I am indebted to 
D.R.T. Jones, D. Kutasov, K. Sibold,
W. Zimmermann, G. Zoupanos and, in particular, to Jisuke Kubo.
It is a pleasure to thank Wolfhart Zimmermann, and the
theory group of the Max Planck Institut f\"ur Physik - Werner Heisenberg
Institut -, for their kind hospitality in M\"unchen.

This work has been supported in part by the National Science Foundation,
grant PHY 91-23780.

\vskip0.5truein

\end{document}